\documentclass[12pt]{article}
\usepackage[margin=0.7in]{geometry}
\usepackage{graphicx}
\usepackage{amsmath}
\usepackage{amsthm}
\usepackage[colorlinks=true, citecolor=blue, urlcolor=blue ]{hyperref}
\usepackage{graphics}
\usepackage{amsfonts}
\usepackage{float}
\usepackage{array}
\usepackage{amssymb}
\usepackage{hyperref}
\usepackage{bm,nicefrac}
\usepackage{caption}
\usepackage{subcaption}
\usepackage{multirow}
\usepackage{authblk}
\usepackage{comment}
\usepackage{hyperref}
\usepackage[toc,page]{appendix}
\title{A Secure Quantum Key Distribution Protocol Using Two-Particle Transmission}
\author{Pratapaditya Bej }
\date{}
\affil{ExamRoom.AI, Konappa Agrahara, Electronic City Phase 1, Bengaluru-560100, India,}
\author{Vinod Jayakeerthi}

\affil{ExamRoom.AI, 1025 Greenwood Boulevard
Suite 401 Lake Mary, Florida 32746, USA}

\begin{document}

\maketitle
\begin{abstract}
The evolution of Quantum Key Distribution (QKD) relies on innovative methods to enhance its security and efficiency. Unextendible Product Bases (UPBs) hold promise in quantum cryptography due to their inherent indistinguishability, yet they are underutilized in QKD protocols. This work introduces a protocol utilizing UPBs to establish quantum keys between distant parties. Specifically, we propose a protocol utilizing a $3\times 3$ tile UPB, where Alice sequentially transmits subsystem states to Bob through quantum channels. The protocol's security is underpinned by the no-cloning theorem, prohibiting the cloning of orthogonal states. We analyze potential attacks, including intercept-resend and detector blinding attacks when quantum channels are noiseless, and discuss the challenges posed by the indistinguishability of our protocol for eavesdroppers, thereby enhancing QKD security.
\end{abstract}
\section{Introduction}

In the landscape of information theory, cryptography plays a vital role in securing data and communications. However, traditional cryptographic systems like RSA, AES face a grave threat from quantum computing due to their reliance on classical computational limitations \cite{shor,grover}. Recognizing the impending threat posed by quantum computing to classical cryptosystems, the imperative for proactive countermeasures becomes evident. One such strategy is the pursuit of post-quantum cryptography \cite{pqc}, which entails the development of novel cryptographic schemes resilient to quantum attacks. However, while post-quantum cryptography offers a partial solution to the problem, it may be susceptible to undiscovered quantum algorithms, leaving its efficacy and long-term security in question. In contrast, Quantum Key Distribution (QKD) stands out as the ultimate solution, leveraging the unbreakable principles of quantum mechanics such as the uncertainty principle and no-cloning theorem \cite{gisin1,scarani0,pirandola1,zurek}.

Quantum cryptography has seen significant advancements in protocol development. In 1984, Bennett and Brassard introduced the pioneering BB84 protocol, which utilizes quantum properties to distribute keys between distant parties securely \cite{bennett0}. Following the introduction of BB84, a series of subsequent protocols surfaced, such as E91 \cite{ekert}, B92 \cite{b92}, BBM92 \cite{bbm92}, and the six-state protocol \cite{bruss1}, significantly broadening the spectrum of quantum secure communication \cite{gisin1,scarani0,pirandola1,lo}. Recent years witnessed the rise of variants such as Device-Independent (DI) QKD \cite{di1,di2}, Measurement-Device Independent (MDI) QKD \cite{mdi1,mdi3}, and Continuous Variable (CV) QKD \cite{cv1,cv3}, driven by both theoretical innovations and experimental implementations \cite{ex2,ex4,ex6,ex8,ex9}.

%Quantum cryptography extends beyond key distribution, encompassing protocols like Quantum Secure Direct Communication (QSDC) \cite{qsdc1,qsdc2}, Quantum Secret Sharing (QSS) \cite{qss1,qss2}, Quantum Identity Authentication (QIA) \cite{qia1}, and Quantum Digital Signature \cite{qds1,qds2}, contributing to the comprehensive landscape of quantum secure communication \cite{gisin1,scarani0,pirandola1}.

Contemporary QKD primarily relies on non-orthogonal states for security. However, the adoption of orthogonal states in cryptographic protocols emerged later, with the inception of the pioneering protocol \cite{or1}. This groundbreaking approach introduced the concept of sending states with controlled time delays, making it nearly impossible for eavesdroppers to intercept an entire state without detection. Moreover, several other studies facilitate the implementation of QKD protocols using orthogonal states, as evidenced by various documented protocols \cite{or2,or3,or4,or6}. Experimental validation has recently been conducted in the case of quantum cryptography based on orthogonal states \cite{or9,or10}. While non-orthogonal state encoding is prevalent, orthogonal state encoding offers potential benefits, such as reduced quantum operation requirements. Understanding the theoretical application of orthogonal states for coding is invaluable due to their innate ability to be distinguished without errors.

The evolution of QKD relies on innovative methods to enhance its security and efficiency. Among these, integrating Unextendible Product Bases (UPBs) may hold promise in quantum cryptography due to their indistinguishability, potentially fortifying the security of quantum communication channels. UPBs are fundamental in quantum information theory \cite{bennett1}. A UPB for a quantum system is an incomplete orthogonal product basis whose complementary subspace cannot be extended to a complete orthogonal basis. UPBs are indistinguishable in the Local Operation and Classical Communication (LOCC) paradigm \cite{bennett1,bennett2,fu,cohen1,rinaldis}.

High-dimensional quantum states offer increased information capacity and noise resilience crucial for securing QKD. Qubit-based systems exhibit a quantum bit error rate threshold of $11\%$, whereas qudit-based protocols show heightened resilience to noise \cite{tittel,cerf,scarani}. The increased noise tolerance also impacts the final secret key rate, as the secret key rate rises with Hilbert space dimensions for a fixed noise level \cite{ekert,liu}. The no-cloning theorem underpins the security of quantum communication, increasing the input state dimension reduces cloning fidelity, emphasizing the benefits of high-dimensional states for quantum cryptography \cite{zurek,navez,bruss}.

No widely recognized protocol frequently incorporates UPBs in QKD protocols. In addressing this gap, we present a protocol that showcases the utilization of UPBs to establish quantum keys between two distant parties. In our protocol, we take the $3\times 3$ tile UPB \cite{bennett1} where Alice sends each subsystem state of a UPB through two quantum channels successively to Bob.  During the transmission of particles, there is a time gap in the particle-sending process such that no eavesdropper has access to two particles simultaneously. The strategy involves dividing the transfer of information into two steps, ensuring that only a portion of the information is transmitted at each step. The security of this approach is guaranteed by the no-cloning theorem concerning orthogonal states \cite{mor}. Throughout the work, we assume noiseless quantum channels, ensuring no information loss during particle transmission through these channels. After Bob receives two particles, a quantum measurement is conducted to distinguish the states. Subsequently, we analyze potential attacks on our protocol, including an efficient intercept-resend attack and detector blinding attacks. We show theoretically that even if an adversary can perfectly blind each single-photon detector, there remains a $50\%$ chance for the eavesdropper to blind the detectors. In our protocol, the indistinguishability of UPBs poses challenges for eavesdroppers attempting to discern between quantum states exchanged during the QKD process, particularly when LOCC is employed by Eve. We also demonstrate that the sequential transmission of the two particles comprising a UPB state through quantum channels hinders the eavesdropper from perfectly distinguishing the transmitted state, even when the eavesdropper employs entanglement as a resource. This inherent indistinguishability of our protocol enhances the security of QKD. The rest of the paper is organized as follows: In Section \ref{s1}, we describe the protocol, and in Section \ref{s2}, we analyze different attacks. Finally, in the last section (Section \ref{s3}), we conclude our results and discuss some future works.

\section{Protocol}\label{s1}

In the context of the quantum key distribution protocol, we focus on a $3\times 3$ dimensional bipartite tiles UPB as outlined in the paper by Bennett et al. \cite{bennett2}. These bases are represented by:

\begin{eqnarray}
\label{eq1}
    |\psi_1\rangle&=&\frac{1}{\sqrt{2}}|0\rangle_{A}(|0\rangle-|1\rangle)_{B}\nonumber\\
    |\psi_2\rangle&=&\frac{1}{\sqrt{2}}(|0\rangle-|1\rangle)_{A}|2\rangle_{B}\nonumber\\
     |\psi_3\rangle&=&\frac{1}{\sqrt{2}}|2\rangle_{A}(|1\rangle-|2\rangle)_{B}\nonumber\\
     |\psi_4\rangle&=&\frac{1}{\sqrt{2}}(|1\rangle-|2\rangle)_{A}|0\rangle_{B}\nonumber\\
     |\psi_5\rangle&=&\frac{1}{3}(|0\rangle+|1\rangle+|2\rangle)_{A}(|0\rangle+|1\rangle+|2\rangle)_{B}
\end{eqnarray}     
  where $A$ and $B$ represent the states of particles $A$ and $B$, respectively. The bases given by Eq.\ref{eq1} are not complete, indicating that $\sum_{i=1}^5|\psi_i\rangle\langle\psi_i|\neq\mathbb{I}$, where $\mathbb{I}$ represents the identity matrix. However, it's important to note that these bases are orthogonal to each other, meaning that $\langle\psi_i|\psi_j\rangle=0$ for all $i\neq j$ within the set $\{1, 2, ..., 5\}$.
 Additionally, they exhibit a degree of nonlocality even in the absence of entanglement. Notably, these bases are indistinguishable under the framework of LOCC \cite {bennett1,bennett2,fu,cohen1,rinaldis,cohen2}.

Therefore, using Eq. \ref{eq1} as the measurement basis, it seems that we cannot form a valid quantum measurement. To construct a valid quantum measurement, we need to ensure that the operators satisfy the completeness relation $\sum_{i=1}^n M_iM_i^{\dagger} = \mathbb{I}$.

To form a quantum measurement along with Eq. \ref{eq1}, we apply the Gram-Schmidt decomposition. First, we begin by constructing four orthogonal states represented as

 $|h_k\rangle$ starting (for $k=6,7,8,9$) from $|\psi_1\rangle$ to $|\psi_4\rangle$.
\begin{eqnarray}
\label{eq2}
 |h_6\rangle=\frac{1}{\sqrt{2}}|0\rangle_{A}(|0\rangle+|1\rangle)_{B}\nonumber\\
    |h_7\rangle=\frac{1}{\sqrt{2}}(|0\rangle+|1\rangle)_{A}|2\rangle_{B}\nonumber\\
     |h_8\rangle=\frac{1}{\sqrt{2}}|2\rangle_{A}(|1\rangle+|2\rangle)_{B}\nonumber\\
     |h_9\rangle=\frac{1}{\sqrt{2}}(|1\rangle+|2\rangle)_{A}|0\rangle_{B}
\end{eqnarray}
i.e. $\langle h_k|\psi_i\rangle=0 \;\;\; \text{for}\;\; i=1,...,4 \;\;\; \text{and}\;\;\; k=6,...,9$. That means $|h_k\rangle$ and $|\psi_i\rangle$ are orthogonal to each other for $i=1,2,..,4$ and $k=6,...,9$.  Next, we formulate:

\begin{eqnarray}
\label{eq3}
     |\psi_6\rangle&=&\alpha_6(|h_6\rangle-\sum_{i=1}^5\langle\psi_i|h_6\rangle|\psi_i\rangle)\nonumber\\
     |\psi_k\rangle&=&\alpha_k(|h_k\rangle-\sum_{i=1}^5\langle\psi_i|h_k\rangle|\psi_i\rangle-\sum_{j=6}^{k-1}\langle\psi_j|h_k\rangle|\psi_j\rangle) \; \;\;\;\text{for}\;\; k=7,8,9
\end{eqnarray}

where $\alpha_k$ is the normalization constant of the $k$-th state that can be easily evaluated, and these values are $\alpha_6=\sqrt{\frac{9}{7}}$, $\alpha_7=\sqrt{\frac{7}{5}}$, $\alpha_8=\sqrt{\frac{5}{3}}$, and $\alpha_9=\sqrt{3}$.

Now, $\sum_{i=1}^9|\psi_i\rangle\langle\psi_i|=\mathbb{I}$ (identity matrix). Therefore, we form a valid quantum measurement by using $|\psi_1\rangle$ to $|\psi_9\rangle$ as the basis. Additionally, it's important to note that each of these basis states is orthogonal to the others, meaning that $\langle\psi_i|\psi_j\rangle=0$ for all $i\neq j$ within the set $\{|\psi_1\rangle,  ...,|\psi_9\rangle\}$.
This specific set of quantum states $|\psi_1\rangle, \ldots, |\psi_5\rangle$ 
 are product bases and $|\psi_6\rangle, \ldots, |\psi_9\rangle$ are entangled bases.

The protocol is as follows:

\textbf{Step 1}: Alice begins by preparing two quantum particles, labeled as A and B, randomly in one of five quantum states using Eq.\ref{eq1}.  
    
\textbf{Step 2:}  After randomly preparing the state in one of the five quantum states using Eq.\ref{eq1}, Alice sends either particle A or B randomly to Bob through quantum channel number 1. Upon receiving the particle, Bob informs Alice through an open classical communication channel.

\textbf{Step 3:} Alice then sends the second particle to Bob through channel 2. Importantly, she only sends the second particle after receiving confirmation that the first particle has reached Bob. This sequential transmission prevents potential eavesdroppers from having simultaneous access to both particles, ensuring security. We assume that both quantum channels are noiseless.

For example, if Alice wants to send the state $|\psi_i\rangle$ to Bob, she may send the second particle $B$ through channel 1 and then $A$ through channel 2, creating the sequence $BA$. Alternatively, she may send particle $A$ first through channel 1 and then $B$ through channel 2, resulting in the sequence $AB$. Alice will keep a record of which path she is sending each particle. In the $AB$ sequence, the two-qutrit state received by Bob will be given by Eq.\ref{eq1}.  In the $BA$ sequence, after the two-qutrit state reaches Bob, he will obtain the state described by Eq.\ref{eq4}, thus forming another tile UPB.

\begin{eqnarray}
\label{eq4}
    |\xi_1\rangle&=&\frac{1}{\sqrt{2}}(|0\rangle-|1\rangle)_{B}|0\rangle_{A}\nonumber\\
    |\xi_2\rangle&=&\frac{1}{\sqrt{2}}|2\rangle_{B}(|0\rangle-|1\rangle)_{A}\nonumber\\
     |\xi_3\rangle&=&\frac{1}{\sqrt{2}}(|1\rangle-|2\rangle)_{B}|2\rangle_{A}\nonumber\\
     |\xi_4\rangle&=&\frac{1}{\sqrt{2}}|0\rangle_{B}(|1\rangle-|2\rangle)_{A}\nonumber\\
     |\xi_5\rangle&=&\frac{1}{3}(|0\rangle+|1\rangle+|2\rangle)_{B}(|0\rangle+|1\rangle+|2\rangle)_{A}
\end{eqnarray}

It is essential to note that $|\xi_i\rangle$ for all $i\in\{1,...,5\}$ are not orthogonal to the set $|\psi_1\rangle, \ldots, |\psi_9\rangle$. Therefore, it is impossible to distinguish all states of Eq.\ref{eq4} using $|\psi_1\rangle, \ldots, |\psi_9\rangle$ as measurement bases.

In the context of Quantum Key Distribution (QKD), the principle of sequential sending, where Alice transmits the second particle to Bob only after receiving confirmation of the safe arrival of the first particle, aligns with the fundamental prerequisites for employing a set of orthogonal product states in five state composite systems where each subsystem are nonidentical and non-orthogonal to each other (for detail see table \ref{table1}). These prerequisites, as established in the QKD scheme, demand that within the density matrix of any subsystem (represented as $\rho_{S|i}$, with $S$ being either subsystem $A$ or $B$), there must exist at least one $\rho_{S|j}$ that differs from $\rho_{S|i}$ and lacks orthogonality with it. This condition, grounded in the laws of quantum mechanics \cite{mor}, upholds the standard no-cloning theorem \cite{zurek}, which is a cornerstone of quantum security. Therefore, the sequential sending of particles, a vital element of the QKD protocol, serves as a practical implementation of the quantum principles that safeguard the security of quantum communication.

\textbf{Step 4:} Upon receiving both particles A and B, Bob performs a collective measurement on them. This measurement is carried out using the basis of the nine quantum bases described earlier ($|\psi_1\rangle$ to $|\psi_9\rangle$) in Eq.\ref{eq1} and Eq.\ref{eq3}. It helps Bob determine the quantum state in which the two-particle system has been prepared.

If Alice sends the $AB$ sequence, Bob can successfully distinguish the state with certainty by forming a quantum measurement with the bases Eq.\ref{eq1} and Eq.\ref{eq3} ($|\psi_1\rangle$,\ldots,$|\psi_9\rangle$). However, when Alice sends the $BA$ sequence to Bob, he will not be able to distinguish the state with certainty, but Bob will register a click on one of his measurement bases. For both sequences, Bob will record all the clicks made by his measurements. Since Alice randomly prepares and sends the state in either the $AB$ or $BA$ sequence, both ways introduce randomness in these two situations; one is during the state preparation process, and the other is during the sequential particle-sending process.

 \textbf{Step 5:} After the measurement process, Bob will communicate with Alice through classical channels to know the sequence of the particle-sending process. If Alice sends the $AB$ sequence, Bob will retain the measurement results and record the information about which state he distinguished. If Alice sends the $BA$ sequence, Bob will discard the results. Alice and Bob have a predefined agreement on how to assign bit values based on the measured quantum states 

 Repeating the entire procedure multiple times, Alice and Bob generate a random bit string. This bit string serves as the raw key for encryption purposes.  

\textbf{Step 6:} To check for potential eavesdropping, Alice and Bob randomly sample and compare bits from their raw key. If the correlations between their bits remain intact with exact values, they can conclude that there is no eavesdropper. If the raw key remains unaltered and secure, Alice and Bob can confidently use the rest of the results as a cryptographic key for secure communication.
 However, if they suspect eavesdropping or find discrepancies during the random bit comparisons, they take security precautions by discarding the entire key and redistributing it.

In the forthcoming section, we delve into a comprehensive analysis of our protocol, shedding light on both the intercept-resend attack and the detector blinding attack, pertinent to our innovative patent-pending device.

\section{Protocol analysis}\label{s2}
Here, we explore two potential security vulnerabilities: the intercept-resend attack and the detector blinding attack. Initially, we delve into the intercept-resend attack, where an eavesdropper intercepts the quantum channel, measures the first particle, and sends a particle to Bob based on this measurement outcome. Subsequently, during the transmission of the second particle, the eavesdropper measures it and sends another particle to Bob based on the measurement results of the first particle. Additionally, we conduct an analysis of the detector blinding attack on the protocol in the subsequent section.

\subsection{Intercept resend attack}
The resend-intercept attack is one of the most significant threats to the security of quantum key distribution (QKD) protocols \cite{ira1}. This attack allows an eavesdropper to intercept and measure the quantum states transmitted by Alice, then resend them to Bob while pretending to be Alice. If Eve is successful, she can obtain the shared secret key between Alice and Bob, compromising the security of their communication. In our protocol, two particles labeled A and B are sent through a quantum channel but with a time delay. Importantly, Eve, the potential eavesdropper, does not have simultaneous access to both particles.
There are two types of resend intercept attack strategies in our protocol.

\textbf{First:} Eve intercepts the first particle sent by Alice through path 1.
She performs a measurement on the particle to determine its quantum state. Based on the measurement outcome, Eve prepares a new particle in the same quantum state.
Eve sends the prepared particle to Bob through path 1, pretending to be Alice. 

Next, Eve intercepts the second particle sent by Alice through path 2.
She analyzes the measurement outcome of the first particle to determine the basis (measurement reference) used by Alice for the second particle. Based on this information, Eve performs a specific measurement on the second particle to determine its quantum state. Using the basis information from the first particle and the new measurement outcome of the second particle, Eve prepares a new particle in the corresponding state.
Eve sends the prepared particle to Bob through path 2, again mimicking Alice.

In our protocol, Alice randomly prepares a UPB state from Eq.\ref{eq1} and sends the particles through two different paths, randomly selecting either path 1 or path 2. If Alice sends the two particles in the $AB$ sequence through these channels, then the state observed by Eve would follow Eq.\ref{eq1}. On the other hand, if Alice sends the $BA$ sequence, then the two-particle state observed by Eve would align with Eq.\ref{eq4}. Eve does not know which specific state is transmitted through these channels, but she is aware that the intercepted state belongs to the set of 10 states $\{|\psi_1\rangle,\ldots,|\psi_5\rangle,|\xi_1\rangle,\ldots,|\xi_5\rangle\}$ represented by Eq.\ref{eq1} and Eq.\ref{eq4}. For Eve, all the 10 states are equally probable. In this particular eavesdropping scenario, Eve employs a sequential approach.

When the first particle is intercepted in path 1, she conducts an orthogonal measurement using the basis $\{|0\rangle\langle0|,   |1\rangle\langle1|, |2\rangle\langle2|\}$.
If she gets the first particle in the state $|0\rangle\langle0|$, she infers the possible two-particle states for A and B, which are $|\psi_1\rangle$, $|\psi_2\rangle$, $|\psi_5\rangle$, $|\xi_1\rangle$, $|\xi_4\rangle$ and $|\xi_5\rangle$ with  1/10, 1/20, 1/30, 1/20, 1/10, 1/30 probabilities of each. Eve sends the first particle, having measured it, to Bob with the results of her measurement. When the second particle arrives, Eve intercepts it as well. She measures the second particle by using the bases $\{|0-1\rangle\langle0-1|, |0+1\rangle\langle0+1|, |2\rangle\langle2|\}$. Based on this measurement, she sends the second particle to Bob. Eve's measurement results on the second particle, particularly if she observes $|0-1\rangle\langle0-1|$, then the two-particle state is either $|\psi_1\rangle$ with 1/10 probability or $|\xi_1\rangle$ with 1/80 probability or $|\xi_4\rangle$ with 1/40 probability, which has collapsed to $|0\rangle|0-1\rangle$. If the second particle is observed in the state $|0+1\rangle\langle0+1|$, the two-particle state collapses to  $|0\rangle|0+1\rangle$. Bob then has a partial probability of 2/270 or 1/80 or 1/40 or 2/270 to find the two-particle state in either $|\psi_5\rangle$ or $|\xi_1\rangle$  or $|\xi_4\rangle$ or $|\xi_5\rangle$ respectively. If the second particle is observed in the state $|2\rangle\langle2|$, the two-particle state collapses to $|0\rangle|2\rangle$. Bob then has a partial probability of 1/40 or 1/270 or 1/20 or 1/270 to find the two-particle state in either $|\psi_2\rangle$ or $|\psi_5\rangle$  or $|\xi_4\rangle$ or $|\xi_5\rangle$ respectively. 
 So, when Eve gets the 1st particle in $|0\rangle$ and after getting the information of the first particle if Bob does a second measurement on the second particle with the measurement $\{|0-1\rangle\langle0-1|, |0+1\rangle\langle0+1|, |2\rangle\langle2|\}$, then the probability for Eve to eavesdrop without detection is $\frac{1}{10}+\frac{1}{80}+\frac{1}{40}+\frac{2}{270}+\frac{1}{80}+\frac{1}{40}+\frac{2}{270}+\frac{1}{40}+\frac{1}{270}+\frac{1}{20}+\frac{1}{270}=0.2722$.

Instead of performing the measurement $\{|0-1\rangle\langle0-1|, |0+1\rangle\langle0+1|, |2\rangle\langle2|\}$ on the second particle, Eve might choose alternative measurements like $\{|0\rangle\langle0|, |1\rangle\langle1|, |2\rangle\langle2|\}$ or  $\{|1-2\rangle\langle1-2|, |1+2\rangle\langle1+2|, |0\rangle\langle0|\}$. However, regardless of the measurement she chooses, the total probability of her successfully eavesdropping without detection remains the same, which is 0.2722, when the first particle is in the state $|0\rangle$.

Similarly in the same way, when Eve gets the 1st particle in $|1\rangle\langle1|$ and after getting the information of the first particle if Bob does a second measurement on the second particle with the measurement $\{|0\rangle\langle0|, |1\rangle\langle1|, |2\rangle\langle2|\}$, then the probability for Eve to eavesdrop without detection is $=0.1222$.

Now, when Eve intercepts the 1st particle in $|2\rangle\langle2|$, and if Bob performs a second measurement on the second particle with the measurement $\{|1-2\rangle\langle1-2|, |1+2\rangle\langle1+2|, |0\rangle\langle0|\}$, then the probability for Eve to eavesdrop without detection is calculated to be 0.2722. Regardless of Eve's choice of alternative measurements, such as $\{|0\rangle\langle0|, |1\rangle\langle1|, |2\rangle\langle2|\}$ or  $\{|0-1\rangle\langle0-1|, |0+1\rangle\langle0+1|, |2\rangle\langle2|\}$, the total probability of her successfully eavesdropping without detection remains the same at 0.2722 when the first particle is in the state $|2\rangle\langle2|$.

So the total probability that Eve eavesdrops on the key information without being detected is $0.2722+0.1222+0.2722=0.6666$.

\textbf{Second:}
The second eavesdropping strategy involves Eve focusing on the second particle, in path 2. Unlike the first strategy, this approach could lead to disruptions in particles in path 2. The motivation for this strategy is driven by the inherent symmetry between particles A and B within our subsystem.
As our UPB state's subsystems are symmetric in this strategy, the resend intercept attack probability remains the same as the initial probability of 0.6666.  
\vspace{0.5cm}

The calculation of the intercept-resend attack is based on LOCC. Cohen's paper \cite{nc2} demonstrated that using LOCC and one ebit of entanglement resource, perfect distinction of the $3\times 3$ UPB is achievable. However, in our scenario, if Eve attempts to distinguish the transmitted UPB states using LOCC with the entanglement resource, perfect distinction isn't possible. This is due to the random sequential transmission of the particles of each UPB state through the quantum channels, which makes her aware that the intercepted state belongs to the set of 10 states $\{|\psi_1\rangle,\ldots,|\psi_5\rangle,|\xi_1\rangle,\ldots,|\xi_5\rangle\}$. Among this set, some states are non-orthogonal, preventing perfect distinguishability. To determine the maximum success probability of unambiguous state discrimination, even if Eve employs two-particle measurements on each UPB state, we utilize a principle outlined in \cite{nc1}. In this scenario, where a quantum system is prepared in one of the $n$ states $|\phi_i\rangle,\ldots,|\phi_n\rangle$ in a $d$-dimensional Hilbert space with probabilities $p_1,\ldots,p_n$, the upper bound for the maximal success probability of unambiguous discrimination among $n$ states using any measurement $\{M_m\}$ is given by:

\begin{equation}
\label{eq5}
D_m(p_1,\ldots,p_n,|\phi_1\rangle,\ldots,|\phi_n\rangle,\{M_m\}) \leq 1 - \frac{1}{(n-1)}\sum_{i \neq j} \sqrt{p_ip_j} |\langle\phi_i|\phi_j\rangle|
\end{equation}

Here, $n$ represents the number of states to be distinguished, $|\phi_i\rangle$ denotes each state to be discriminated, and $p_i$ indicates the prior probability of the $|\phi_i\rangle$ state. In our specific system, if Eve aims to discriminate the transmitted state unambiguously among the set $\{|\psi_1\rangle,\ldots,|\psi_5\rangle,|\xi_1\rangle,\ldots,|\xi_5\rangle\}$ where each state is equally probable, the maximum success probability would be $\frac{8}{9}$. Thus the perfect discrimination of state in this set is not possible.

\subsection{Detector blinding attack:}

A detector blinding attack targets quantum key distribution (QKD) protocols, exploiting vulnerabilities in quantum signal detection to gain unauthorized access to sensitive information without raising the quantum bit error rate (QBER). In protocols like BB84, Eve intercepts Alice's qubits, measures them, and blinds Bob's detectors with intense light, controlling their firing to match her measurements. Success relies on manipulating Bob's random number generator, rendering detectors blind. Without control, blinding alone wouldn't reveal the key, making detection challenging. Thus, the attack primarily hinges on controlling Bob's base selection, highlighting the importance of random number generator security in QKD \cite{db1}.

Our protocol needs a two-particle joint measurement to distinguish the state, Eve cannot access both particles simultaneously during transmission. Hence, for eavesdropping, Eve needs to perform a single-particle measurement on the first particle and then choose the second-particle measurement based on the outcome of the first measurement. We assume that Eve can successfully blind the single-photon detectors.

In a linear optics setup, if Eve employs a faked-state attack, she intercepts the first particle, causing detector clicks based on her chosen measurement basis. Subsequently, based on the outcome of the first measurement, she selects which state to fake from the set $\{|\psi_1\rangle,\ldots,|\psi_5\rangle,|\xi_1\rangle,\ldots,|\xi_5\rangle\}$. She then sends the first particle's polarization state to Bob, accompanied by intense circularly polarized light. After intercepting the second particle sent by Alice, Eve performs a specific measurement on it to determine its quantum state. Depending on the predetermined state she intends to send after the first measurement, Eve forwards the polarization state of the second particle to Bob, again accompanied by intense circularly polarized light. Eve manipulates Bob's detectors to fire as she desires, aligning with her measurement results.

In our protocol, Alice randomly prepares and sends particles through quantum channels. After the measurement, Bob receives confirmation via a classical channel regarding the sequence ($AB$ or $BA$) in which Alice sent the state. During this confirmation process, Bob can detect the presence of an eavesdropper. For instance, if Alice sends an $AB$ sequence state but Eve decides to send a $BA$ sequence state based on the first measurement outcome, Bob's detectors will not click any of the $AB$ sequence states (Eq.\ref{eq1}) but instead detect a different sequence, indicating interference. Therefore, Bob becomes aware of the eavesdropper. However, Eve has a $50\%$ chance of blinding Bob's detectors successfully since her decision will align with either the $AB$ or $BA$ sequence half of the time.

As an example, let's consider Alice sending a state $|\psi_1\rangle$ in the AB sequence. Eve intercepts the first particle and measures it in the $\{|0\rangle\langle0|,|1\rangle\langle1|,|2\rangle\langle2|\}$ bases, obtaining a click in the $|0\rangle\langle0|$ basis. Based on this measurement outcome, Eve infers that the two-particle state could be one among the set $\{|\psi_1\rangle, |\psi_2\rangle, |\psi_5\rangle, |\xi_1\rangle, |\xi_4\rangle,|\xi_5\rangle\}$. Suppose Eve decides to send a fake state $|\xi_1\rangle$ and transmits the subsystem states to Bob sequentially. Due to the fake state sent by Eve, Bob's detector will not click any of the AB sequence states (Eq.\ref{eq1}). When Bob communicates with Alice to confirm the sequence via the classical channel, he realizes that eavesdropping has occurred, even if Eve can intercept and listen to the classical messages.

 Lastly, the security amplification process is also present, where Alice and Bob will compare their certain shared bits, right or wrong, during that process to identify the attack.

\section{Conclusion:}\label{s3}

A primary contribution of our research lies in the investigation of orthogonal state encoding and the incorporation of UPB into QKD protocols. Through theoretical analysis, we have demonstrated the viability of employing $3\times 3$ tile UPB \cite{bennett1,bennett2} to establish secure quantum communication channels between distant parties.

Furthermore, our study has provided insights into the vulnerabilities of QKD protocols through the analysis of intercept-resend \cite{ira1} and detector blinding attacks \cite{db1}. An intercept-resend attack is a cybernetic attack on quantum key distribution systems, where the attacker, often referred to as Eve, intercepts quantum signals intended for the recipient, Bob, without being detected. In the BB84 protocol, there is a $75\%$ chance for an eavesdropper to remain undetected. By quantifying the probabilities and outcomes associated with these attacks within the framework of our protocol, we have identified an efficient intercept-resend attack, resulting in a $66\%$ chance of successful eavesdropping without detection. There is a paper \cite{nc2} that demonstrated how one ebit of entanglement enables perfect discrimination of $3\times 3$ UPB states via LOCC. However, our protocol incorporates sequential particle transmission, complicating the eavesdropper's ability to distinguish the UPB state. Upon observing the transmitted UPB state, the eavesdropper encounters a set $\{|\psi_1\rangle,\ldots,|\psi_5\rangle,|\xi_1\rangle,\ldots,|\xi_5\rangle\}$ containing non-orthogonal states. This non-orthogonality hinders perfect state discrimination, resulting in imperfect discrimination even if she uses entanglement. Furthermore, we showed that even if an eavesdropper employs any measurement, she would only be able to unambiguously distinguish the transmitted state with a maximum success probability of $\frac{8}{9}$.

Other eavesdropping strategies include Eve storing the first particle in a quantum memory and sending a random state to Bob, followed by a joint measurement on both particles. However, success in this scenario is limited to $\frac{1}{3}$. In orthogonal state quantum cryptography, each subsystem of the orthogonal bipartite state is successively sent through the two quantum channels, dividing information transmission into two stages. This ensures that only a fraction of information is conveyed at any given moment. Additionally, quantum memories have limited storage time, and once this time limit is reached, extracting information becomes impossible \cite{qm1,qm2,qm3,qm4,qm5}. Furthermore, the delay time in our scenario exceeds the time taken by particles to travel from Alice to Bob, enhancing security due to the no-cloning theorem applicable to orthogonal states \cite{mor}.

While quantum key distribution (QKD) protocols are renowned for their unconditional secrecy, the security of QKD hardware hinges significantly on the intricacies of its implementation. A detector blinding attack poses a security threat in quantum cryptography, allowing an eavesdropper to manipulate quantum detectors to intercept communication without detection \cite{db1}. Even if we consider the possibility of an adversary successfully blinding the single-photon detectors, our protocol has a $50\%$ chance for an eavesdropper to achieve successful detector blinding. It's worth noting that our protocol involves the transmission of two-particle states, which are product states.

Recent advancements in hardware and protocols offer robust countermeasures against detector-blinding attacks \cite{db2,db3,db5}. Enhancing the integrity of Bob's detectors is crucial, potentially achieved through incorporating randomized control mechanisms into their operation. By introducing randomness, it becomes more challenging for adversaries to predict and manipulate the detector's behavior, thus bolstering its resilience against manipulation attempts.

Our investigation into utilizing $3\times 3$ tile UPBs for establishing secure quantum keys highlights their indistinguishability under the LOCC paradigm. While our protocol restricts eavesdroppers' access to two-particle states, a time gap between transmissions limits adversaries to LOCC tools for eavesdropping. We acknowledge the potential existence of better protocols for utilizing UPBs in QKD and generalizing them for $d\times d$ dimensions. To extend our protocol, one may follow the same steps and formulate a quantum measurement. Let $|\psi_1\rangle,\ldots,|\psi_l\rangle$ form the UPB in $\mathbb{C}^d\otimes \mathbb{C}^d$ \cite{upb1,upb2}, where $\sum_{i=1}^l|\psi_i\rangle\langle\psi_i|\neq\mathbb{I}$ and $|\psi_l\rangle$ is the stopper basis. Construct complete measurement bases by first forming orthogonal bases $|h_{l+1}\rangle,\ldots,|h_{d^2}\rangle$ orthogonal to each basis of $|\psi_1\rangle,\ldots,|\psi_{l-1}\rangle$, with $\langle h_k|\psi_i\rangle=0$ for $i=1,\dots,l-1$ and $k=l+1,\ldots,d^2$ and then, formulate:
\begin{eqnarray}
\label{eq6}
|\psi_{l+1}\rangle&=& \alpha_{l+1}\Big(|h_{l+1}\rangle-\sum_{i=1}^l\langle\psi_i|h_{l+1}\rangle|\psi_i\rangle\Big)\nonumber\\
|\psi_{k}\rangle&=& \alpha_{k}\Big(|h_{k}\rangle-\sum_{i=1}^l\langle\psi_i|h_{k}\rangle|\psi_i\rangle-\sum_{j=l+1}^{k-1}\langle\psi_j|h_{k}\rangle|\psi_j\rangle\Big)  \; \;\;\;\text{for}\;\; k=l+2,\ldots,d^2
\end{eqnarray}
such that $\sum_{i=1}^{d^2}|\psi_i\rangle\langle\psi_i|=\mathbb{I}$. Here, $\alpha_k$ is the normalization constant of the state $|\psi_k\rangle$. Let's say, if the number of UPB in a $d\times d$ system is $d^2-2d+1$, then one has to form $2d-1$ number of entangled bases to construct a complete set of measurement bases \cite{upb1}. Using our protocol, one can distribute the quantum key between distant parties using the $d\times d$ UPBs. Exploring the optimal intercept resend attack probability would be a compelling avenue for future research.

In conclusion, our study advances quantum cryptography by elucidating orthogonal state encoding principles, UPBs, and their role in fortifying quantum communication. Rigorous analysis of intercept-resend and detector blinding attacks provides valuable insights for establishing secure communication channels in the quantum technology era.

\section*{Acknowledgements:} 
This work is for the organization \href{https://examroom.ai/}{ExamRoom.AI.} The authors are also thankful to Priti Kumari, and Ritobroto Mohanta for the helpful discussions.

\vspace{14cm}
\appendix{\textbf{\textit{{\Large Appendix:}}}}
\section{Reduced states:}

The reduced subsystems of Eq.\ref{eq1} and Eq.\ref{eq3} for A and B are evaluated by using the relation $\rho_{A(B)|i}=\text{Tr}_{B(A)}(|\psi_i\rangle_{AB}\langle\psi_i|)$, where $\rho_{A(B)}$ represents the reduced state of A and B for the $i$-th state $|\psi_i\rangle$.  
\begin{center}
\begin{table}[H]
\caption{Reduced density matrices of individual subsystems}
\label{table1}
\begin{tabular}{ |m{9cm}|m{9cm}| } 

 \hline
 $\rho_{A|i}$ &  $\rho_{B|i}$ \\
 \hline
  $\rho_{A|1}=|0\rangle\langle0|$ &  $\rho_{B|1}=\frac{1}{2}|0-1\rangle\langle0-1|$ \\
  \hline
$\rho_{A|2}=\frac{1}{2}|0-1\rangle\langle0-1|$ &  $\rho_{B|2}=|2\rangle\langle2|$ \\
  \hline
  $\rho_{A|3}=|2\rangle\langle2|$ &  $\rho_{B|3}=\frac{1}{2}|1-2\rangle\langle1-2|$ \\
  \hline
  $\rho_{A|4}=\frac{1}{2}|1-2\rangle\langle1-2|$ &  $\rho_{B|4}=|0\rangle\langle0|$ \\
  \hline
  $\rho_{A|5}=\frac{1}{3}|0+1+2\rangle\langle0+1+2|$ &   $\rho_{B|5}=\frac{1}{3}|0+1+2\rangle\langle0+1+2|$ \\
  \hline
  $\rho_{A|6}=\frac{1}{21}(4|0-1\rangle\langle0-1|+4|0-2\rangle\langle0-2|-2|1-2\rangle\langle1-2|+9|0\rangle\langle0|)$ &  $\rho_{B|6}=\frac{1}{42}(19|0+1\rangle\langle0+1|+2|0-2\rangle\langle0-2|+2|1-2\rangle\langle1-2|-2|0\rangle\langle0|-2|1\rangle\langle1|)$ \\
  \hline
 $\rho_{A|7}=\frac{1}{70}(25|0+1\rangle\langle0+1|+2|1-2\rangle\langle1-2|+10|0-2\rangle\langle0-2|-10|0\rangle\langle0|+6|1\rangle\langle1|)$ &  $\rho_{B|7}=\frac{1}{35}(4|0+1\rangle\langle0+1|+3|0-2\rangle\langle0-2|+3|1-2\rangle\langle1-2|-3|0\rangle\langle0|-3|1\rangle\langle1|+21|2\rangle\langle2|)$ \\
  \hline
  $\rho_{A|8}=\frac{1}{15}(|1-2\rangle\langle1-2|+3|1\rangle\langle1|+10|2\rangle\langle2|)$ &  $\rho_{B|8}=\frac{1}{30}(9|1+2\rangle\langle1+2|+2|0-1\rangle\langle0-1|+6|0-2\rangle\langle0-2|+2|1\rangle\langle1|-6|2\rangle\langle2|)$ \\
  \hline
  $\rho_{A|9}=\frac{1}{6}|1+2\rangle\langle1+2|+\frac{4}{6}|1\rangle\langle1|$ &  $\rho_{B|9}=\frac{1}{3}|0-1\rangle\langle0-1|+\frac{1}{3}|1\rangle\langle1|$ \\
  \hline
\end{tabular}
\end{table}
\end{center}

\end{document}